\documentclass[a4paper,fleqn]{cas-sc}

\usepackage[authoryear]{natbib}
\usepackage{amsmath,amssymb,amsfonts}
\usepackage{algorithmic}
\usepackage{graphicx}
\usepackage{textcomp}
\usepackage{tabularx}

\usepackage{enumitem}
\usepackage{multirow}
\usepackage{bm}

\def\tsc#1{\csdef{#1}{\textsc{\lowercase{#1}}\xspace}}
\tsc{WGM}
\tsc{QE}

\begin{document}
\let\WriteBookmarks\relax
\def\floatpagepagefraction{1}
\def\textpagefraction{.001}

\shorttitle{}    

\shortauthors{}  

\title [mode = title]{Situational Awareness as the Imperative Capability for Disaster Resilience in the Era of Complex Hazards and Artificial Intelligence}  



%

\author[1]{Hongrak Pak}[orcid=0000-0003-3899-1156]

\cormark[1]


\ead{hongrak822@tamu.edu}


\credit{Conceptualization, Data curation, Investigation, Methodology, Writing – original draft, Writing – review \& editing}

\affiliation[1]{organization={UrbanResilience.AI Lab, Zachry Department of Civil and Environmental Engineering, Texas A\&M University},
            city={College Station},
            postcode={77843}, 
            state={TX},
            country={USA}}

\author[1]{Ali Mostafavi}
\credit{Conceptualization, Data curation, Investigation, Methodology, Writing – original draft, Writing – review \& editing}




\cortext[1]{Corresponding author}



\begin{abstract}
Disasters frequently exceed established hazard models, revealing “blind spots” where unforeseen impacts and vulnerabilities hamper effective response. This perspective paper contends that situational awareness (SA)—the ability to perceive, interpret, and project dynamic crisis conditions—is an often overlooked yet vital capability for disaster resilience. While risk mitigation measures can reduce known threats, not all hazards can be neutralized; truly adaptive resilience hinges on whether organizations rapidly detect emerging failures, reconcile diverse data sources, and direct interventions where they matter most. We present a technology–process–people roadmap, demonstrating how real-time hazard nowcasting, interoperable workflows, and empowered teams collectively transform raw data into actionable insight. A system-of-systems approach enables federated data ownership and modular analytics, so multiple agencies can share timely updates without sacrificing their distinct operational models. Equally crucial, structured sense-making routines and cognitive load safeguards help humans remain effective decision-makers amid data abundance. By framing SA as a socio-technical linchpin rather than a peripheral add-on, this paper spotlights the urgency of elevating SA to a core disaster resilience objective. We conclude with recommendations for further research—developing SA metrics, designing trustworthy human–AI collaboration, and strengthening inclusive data governance—to ensure that communities are equipped to cope with both expected and unexpected crises. \nocite{*}
\end{abstract}


\begin{highlights}
\item Situational awareness reframes as a core disaster resilience capability.
\item The proposed roadmap integrates technology, process, and people for adaptive disaster response.
\item System-of-systems approach enables federated, interoperable crisis intelligence.
\item SA metrics, human–AI collaboration, and inclusive governance guide future research.
\end{highlights}


\maketitle

\section{Introduction} \label{sec:introduction}
Disasters inevitably bring unanticipated consequences that go beyond the scope of the most advanced and well-organized mitigation or preparedness strategies \citep{omitaomu2010framework, riddell2019exploratory, songsore2017complex, fan2021disaster}. While risk-based planning emphasizes identified hazard profiles and standardized protective measures, actual disaster-induced emergencies are often different from anticipated scenarios. Such deviations reveal critical ‘blind spots’ where responders face a lack of clarity on real-time conditions, urgent intervention points, and immediate response requirements. In this context, situational awareness (SA) is the linchpin that lights up the blind spots and addresses the information scarcity by enabling rapid detection, collective sense-making, and targeted responses to whatever truly unfolds \citep{laurila2023protocol}.
Despite substantial progress in disaster resilience, ranging from advanced hazard modeling to comprehensive preparedness strategies, SA remains comparatively underexplored \citep{o2020situation}. Post-event analyses frequently cite delayed impact recognition, fragmented or contradictory information, and lengthy decision flows as primary obstacles for proactive responses to severe disruptions and cascading failures \citep{hernantes2013learning}. These challenges, however, often stem from deficiencies in SA rather than those in hazard mitigation plans or measures. Furthermore, existing research does partially capture the socio-technical complexities of SA, particularly the distributed decision-making structures and the potential of emerging data sources (e.g., real-time monitoring \citep{middleton2013real, fan2020hybrid, pak2025real}, AI-based damage assessment \citep{cheng2023probabilistic, kamari2022ai, esparza2025ai}) that can enrich awareness if effectively integrated. Such perspectives underscore that SA should be recognized as a fundamental element of resilience frameworks, particularly given that not all disasters can be prevented or mitigated in advance. An effective SA capability forms a dynamic feedback loop wherein communities and responders continuously monitor conditions, interpret signals, and anticipate near-term consequences \citep{endsley2017toward}. When these components work together in a cohesive way against unmanaged threats, rapid responses and decision-making can be made. By minimizing the time lag between a disaster’s onset and the establishment of a common operating picture (COP), ranging from a shared, multi-agency understanding of reality, SA significantly reduces human suffering and infrastructure damage \citep{chen2008coordination, curnin2015role, militello2005large, li2022location}.

Unanticipated and unmitigated impacts represent some of the most hazardous dimensions of disaster scenarios, often referred to as ‘black swan’ events \citep{nafday2009strategies}. Even though they occur occasionally, their impact and consequences fall largely outside routine hazard projections. Traditional risk assessments tend to overlook these rare or compound crisis scenarios, instead focusing on well-defined hazards. In contrast, SA provides the necessary agility and intelligence to detect and respond to black swan events before they metastasize. By integrating diverse data streams, such as sensor outputs, social media analytics, and infrastructure monitoring, SA platforms can spot subtle anomalies or interdependencies that are typically overlooked by most conventional approaches. This ongoing vigilance makes it possible to identify weak signals of an impending shock, whether it be an emergent pandemic, a sudden infrastructure failure, or an atypical cascade of weather extremes. In turn, SA not only delivers immediate clarity for decision makers but also fosters a proactive monitoring culture. Once anomalies are identified, response organizations are better equipped to act decisively and pivot swiftly, rather than scrambling after a crisis has already engulfed communities. Thus, SA enables authorities to reinforce critical lifelines, preposition essential resources, and issue well-targeted public warnings. Through such anticipatory actions, SA shifts disaster management from reactive crisis response to proactive adaptation, thereby reducing the potential severity of black swan disruptions.
A primary rationale for enhancing SA within disaster management lies in the urgent need to shift away from reactive patterns that still dominate during many disasters. Despite advances in risk assessments and hazard forecasts, far too often, communities and agencies discover critical threats only when they have already begun to intensify. SA, by contrast, enables a proactive stance: it continuously updates our understanding of evolving hazards, aligns disparate data sources into a unified picture, and flags early warning signals of emerging crises. Through this real-time SA intelligence, decision makers can deploy resources, issue warnings, or re-route supply chains before conditions deteriorate, thereby narrowing the blind spots in which a crisis can escalate uncontrollably.
Fundamentally, SA improves the ‘waiting for the worst’ manner in traditional approaches. It allows responders to detect imminent threats and anticipate secondary consequences such as cascading infrastructure failure or sudden population movements. When integrated effectively into operational frameworks, SA redefines disaster management from a scrambled and last-minute reaction into a deliberate and forward-leaning strategy that hardens defenses and minimizes harm \citep{fan2020social, farahmand2022network, yuan2022smart}. Moreover, SA underpins the entire disaster preparedness cycle, including ‘blue-sky’ and ‘grey-sky’ phases. During a routine, non-crisis period (‘blue-sky’ phase), it aids in identifying latent vulnerabilities and informing preemptive mitigation strategies. In the ‘grey-sky’ phase, as hazards become imminent or unfold, SA drives agile decision-making and resource deployment, limiting preventable losses \citep{huang2015geographic, harrald2007shared}. In the post-disaster recovery phase, SA supports effective rehabilitation efforts and maintains an accurate current understanding of restoration priorities and progress \citep{brown2008supporting, inspector2017review, coleman2023lifestyle}. However, realizing the full potential of SA is inherently challenging. Disaster environments are characterized by fragmented information flows, organizational silos, and high operational pressures, all of which demand seamless coordination among personnel, processes, and technologies. Addressing these complexities is essential to fully integrate SA into resilient disaster management practices.
In the sections that follow, we present a threefold roadmap for strengthening SA as a focal resilience capability. We begin by reviewing the theoretical foundations spanning Endsley’s three-level model and emerging distributed SA frameworks, then illustrate how ‘blue-sky’ and ‘grey-sky’ SA differ in practice. Building on that, we outline key technologies (e.g., hazard nowcasting, infrastructure interdependency mapping), discuss critical processes (data-sharing protocols, integrated workflows), and emphasize the people dimension (collective sense-making, cognitive load management) needed to transform raw data into coordinated, life-saving action. By addressing these interconnected challenges, the field can significantly reduce the current deficiencies surrounding SA. This advancement will ensure that resilience strategies extend beyond mitigating anticipated hazards, enabling a more robust and adaptive response to the unforeseen and complex disruptions that inevitably accompany major disaster events.
Another compelling reason that SA must be elevated is the inherent limitations and uncertainties of risk forecasting. Despite increasingly sophisticated models, forecasting the precise scale, intensity, or course of a hazard remains a complex task—one often confounded by incomplete data and unforeseen interactions. Real‑world crises frequently exceed or diverge from predicted scenarios, leaving purely forecast‑based preparedness vulnerable to blind spots. SA, by contrast, offers a dynamic lens through which decision‑makers can continuously reconcile forecast outputs with actual unfolding conditions. This real‑time intelligence prevents agencies from becoming overdependent on static projections and allows them to calibrate actions as new evidence surfaces, ensuring that response and resource allocation keep pace with the reality on the ground rather than a rapidly outdated forecast.

\section{Theoretical frameworks of situational awareness (SA)}
\subsection{Situational awareness (SA) in high-risk domains and its relevance to disasters}
SA originally emerged as a key concept in high-risk fields such as aviation and military operations, where real-time understanding of a dynamic environment can mean the difference between success and failure under extreme pressure. Seminal work by \citet{endsley2017toward} defines SA as comprising three hierarchical levels: Level 1-Perception of relevant cues, Level 2-Comprehension of their meaning, and Level 3-Projection of future states or events. Although this formulation initially addressed the cognitive processes of a single pilot or operator, it has proven extremely applicable to disaster resilience, where rapidly changing hazard conditions demand a similar cycle of perceiving, interpreting, and anticipating potential impacts.
In a disaster context, Endsley’s model translates to gathering timely hazard and infrastructure data (Level 1: Perception), interpreting how these data affect populations and critical systems (Level 2: Comprehension), and predicting cascading effects (Level 3: Projection), for example, identifying how damaged roads may hinder rescue efforts or how rising floodwaters could compromise a nearby hospital. Because disasters often involve compressed decision timelines and ambiguous signals, each level in SA must operate at pace and scale beyond typical day-to-day emergency operations. Any gap, such as failing to perceive that floodwaters have overtopped critical levees, can undermine both immediate response and longer-term resilience.

\begin{table*}[!htbp]
\caption{Summary of situational awareness (SA) models and frameworks} \label{table:SA summary}
\setlength{\tabcolsep}{3pt}
\begin{tabularx}{\textwidth}{|p{3cm}|p{3cm}|X|} \hline
Framework/Model & Authors (Year) & Main contribution \\ \hline
Three-Level SA model (Individual SA) & \citet{endsley1995measurement} & 
Foundational Model of Situational Analysis (SA) 
\begin{itemize}[leftmargin=*, nosep]
    \item Defines SA as perception, comprehension, and projection.
    \item Basis for most SA research, including disaster contexts.
\end{itemize} 
\\ \hline 
Distributed SA (DSA) Theory & \citet{stanton2006distributed, stanton2016distributed} & DSA Conceptualization
\begin{itemize}[leftmargin=*, nosep] 
    \item SA emerges from distributed systems, not a single person's knowledge.
    \item In complex operations, SA is collectively held by responders, tools, and environment.
    \item Emphasizes information sharing and interaction.
    \item Inspired methods to measure team SA and design systems for COP.
\end{itemize}
 \\ \hline 
Shared SA in Emergency Response & \citet{seppanen2015shared} & Empirical Framework for Multi-Agency Shared Awareness
\begin{itemize}[leftmargin=*, nosep]
    \item Defines critical information requirements, communication processes, and trust components.
    \item Guides agencies in improving information flow and cooperation.
\end{itemize}
 \\ \hline 
Multi-Agency SA Models & \citet{o2020situation} & Review of Safety in Multi-Agency Disaster Response
\begin{itemize}[leftmargin=*, nosep]
    \item Catalogues existing Safety Analysis (SA) models and methods.
    \item Examines SA breakdowns in interagency settings.
    \item Integrates human factors models with disaster case studies.
    \item Proposes improvements like better training for shared SA and interface design for joint teams.
\end{itemize}
 \\ \hline 
Shared SA Formation Framework & \citet{laurila2023protocol} & Systemic Framework for Disaster Scenario Analysis
\begin{itemize}[leftmargin=*, nosep]
    \item Developed for multi-agency disaster simulations.
    \item Maps information gathering, alignment, and use.
    \item Importance of shared Safeguarding (SA) in disaster preparedness.
    \item Measures shared SA's impact on decision-making during exercises.
\end{itemize}
\\ \hline 
Social Media–Based SA Framework & \citet{freitas2020conceptual} & Social Media Decision-Making Framework for Emergency Management
\begin{itemize}[leftmargin=*, nosep]
    \item Utilizes crowdsourced information for emergency management.
    \item Tested on 2017 earthquake and wildfire data.
    \item Found small tweets (about 2\%) to improve responders' safety.
    \item Integrates non-traditional data sources into safety models.
\end{itemize}
 \\ \hline 
\end{tabularx}
\end{table*}

\subsection{From individual to distributed situational awareness (DSA)}
While Endsley’s model continues to serve as a foundational framework, subsequent research highlights that disaster responses extend beyond individual cognition. Multiple agencies (e.g., fire services, law enforcement, EMS, public health, and NGOs), along with technological systems (e.g., sensor networks, GIS platforms, and AI-driven models), collaboratively shape evolving situations. This shift has fueled the concept of Distributed Situational Awareness (DSA): SA emerges from the dynamic interaction of people, technologies, and protocols in real-time, rather than residing in any one head or command post. Scholars such as \citet{stanton2006distributed}, \citet{stanton2016distributed}, and \citet{o2020situation} emphasize that no single vantage point or dataset can entirely capture crisis complexity. Each stakeholder holds partial knowledge, and coherent, system-wide awareness depends on continuous information exchange, cross-verification, and alignment.
In practice, DSA manifests when insights from various sources integrate into a COP. For example, a GIS specialist may detect rising water levels near a substation, field crews may report road washouts, and a regional EOC may monitor shelter capacities. SA becomes truly distributed when these insights synchronize, minimizing blind spots and supporting informed decision-making. Even if each sector knows its domain, SA becomes “distributed” only when those pieces synchronize into a collective model.

\subsection{The OODA Loop: Observe, Orient, Decide, and Act}
A complementary framework to SA is the OODA Loop, developed by \citet{richards2020boyd} for military strategy. It consists of four iterative stages: Observe, Orient, Decide, and Act. In disaster management, “Observe” and “Orient” strongly correlate with Endsley’s Level 1 (Perception) and Level 2 (Comprehension), where teams gather relevant cues (e.g., sensor data, damage reports) and contextualize them within the crisis context. “Decide” and “Act,” meanwhile, map to deploying resources or issuing warnings. The OODA Loop particularly highlights speed and continuous updating: whichever team can process situational cues more rapidly is better positioned to prevent cascading failures. For disasters that escalate hour by hour, such as flash floods or urban fires, the ability to quickly reorient and pivot decisions can drastically reduce losses.

\subsection{Role of international and national frameworks}
In recent years, disaster resilience standards and frameworks increasingly recognize SA as essential for effective response and recovery, more than an operational convenience. \citet{iso22320_2018}, for instance, mandates consistent information flow, shared situational pictures, and interoperability to coordinate multi-agency activities. Similarly, the U.S. NIST Disaster Resilience Framework \citep{nist2015_disaster_resilience_framework} designates SA as a critical outcome of reliable communications, urging all levels of government to maintain a cohesive COP to minimize confusion. At the global level beyond national structures, the United Nations Office for Disaster Risk Reduction (UNDRR) \citep{united2023global} emphasizes that early warning systems must incorporate near real-time intelligence and validated data to guide policy and community action. 
These frameworks often distinguish between "blue sky" and "grey sky" awareness. "Blue sky" refers to ongoing risk monitoring, such as weather forecasts, seismic activity, and baseline mapping of critical infrastructure. It supports community preparedness during normal conditions. In contrast, "grey sky" SA emerges during imminent or active hazards and relies on rapidly updated intelligence. This information underpins tactical decisions, including evacuation routes, resource allocation, and immediate life safety measures.

\subsection{Operational taxonomies and coping with evolving hazards}
Agencies such as the US Department of Homeland Security (DHS) and the Federal Emergency Management Agency (FEMA) further classify SA into operational taxonomies. They specify data categories necessary for a COP, including hazard data, infrastructure status, population vulnerability, etc. By distinguishing “blue sky” data (long-term vulnerability scans) from “grey sky” intelligence (live sensor readings, social media feeds), these taxonomies highlight how SA requirements evolve throughout a disaster. While pre-event monitoring might focus on subtle risk indicators or seasonal patterns, the active response phase demands faster updates and reliable mechanisms to validate and disseminate information across all operational levels, from field teams to national coordination centers. Despite widespread recognition of SA in disaster management frameworks, practical integration remains incomplete. Post-incident reviews often reveal persistent issues, including confusion, inconsistent data sharing, and misaligned assumptions, all of which hinder effective collaboration. These challenges indicate that the process of building and sustaining multi-stakeholder SA is still underdeveloped. Addressing this gap requires leveraging new technologies, creating interoperable processes, and fostering people-centered solutions that embed DSA principles into routine operations.

\section{Blue-sky and Gray-sky situational awareness (SA)}
In practice, SA spans a continuum from steady‑state vigilance before disasters (“blue sky”) to the intense, rapid‑fire context of an active crisis (“grey sky”). While previous sections addressed overarching frameworks like Endsley’s model \citep{endsley1995measurement} and DSA \citep{stanton2006distributed}, the blue sky or grey sky distinction sharpens our understanding of when different SA tasks occur and which stakeholders must engage. Together, these phases encapsulate how communities prepare for creeping vulnerabilities and how they respond when hazards strike with real‑time unpredictability.

\subsection{Blue-sky SA: Proactive and ongoing preparedness}
During normal situations, blue-sky SA includes constantly monitoring and anticipating potential threats. This goes beyond conventional risk assessments and includes mapping critical systems, analyzing redundancies, and identifying infrastructure that might be a single point of failure. By doing so, utilities or local governments can implement minor, progressive safeguards well before an earthquake or storm is expected.

"Steady-state" data streams are also useful for proactive preparedness: meteorological predictions, hydrological signals, and vulnerability indexes can detect little but major changes in risk levels. Early identification of these signals enables organizations to pre-position resources, such as medical supplies, emergency personnel, and sandbags, or to allocate finances to address potential risks, such as fixing backup generators at important shelters. This is compatible with Endsley's model's Level 3 "projection" component, which claims that authorities learn to predict short-term difficulties and act quickly even when there is no imminent threat. Blue-sky SA is, thus, supported by procedures (such as frequent scenario planning exercises) and technology (such as real-time SCADA dashboards), which provide both raw situational data and organized tools for understanding it. Cross-sector collaboration on the human side ensures that no one firm is completely aware of hazards. Cross-sector collaboration prevents a single organization from monopolizing risk information.

\subsection{Grey-Sky SA: Immediate response and adaptive recovery}
Once a hazard looms (or materializes without warning), grey sky situational awareness becomes imperative to guide agile response. Here, the challenge escalates: responders must continuously gauge which roads are blocked, which shelters are full, where injuries are concentrated, and how cascading failures (e.g., power outages) might worsen other hazards. Real‑time data feeds—from drones, IoT sensors, or crowdsourced reports—update the common operating picture by the minute, helping incident commanders decide when to issue evacuation orders, where to dispatch rescue teams, or how to reroute supplies around blocked highways. This dynamic, high‑tempo environment matches DSA precepts, where multiple agencies connect fragments of insight to create a coherent big picture.

“Nowcasting” for hazards such as floods or wildfires illustrates grey sky situational awareness in practice. By monitoring fresh sensor data or meteorological variations, sophisticated models update the crisis prognosis hourly or more frequently—significantly surpassing the conventional daily danger projection. Expeditious feedback mechanisms—such as those exemplified by the OODA Loop—are essential: as circumstances evolve, decision-makers must recalibrate and adjust, frequently resorting to improvisation if conventional strategies are insufficient. Conversely, neglecting to adapt to the hazard's evolution results in blind spots, redundancy, and delayed assistance, each exacerbating the distress of affected communities. Recovery necessitates ongoing situational awareness. As the threat diminishes, agencies shift from immediate assistance to the stabilization and reconstruction of communities. Near-real-time data—such as UAV evaluations of residual debris or mobile phone analytics of population returns—facilitates the allocation of resources to the most vulnerable or neglected communities, ensuring equitable support for all groups. The effective grey sky SA persists until the community's fundamental operations are reinstated. Table~\ref{table:blue-gray} delineates the distinctions in attention, information flows, and operational rhythms between blue sky and grey sky situational awareness. The primary conclusion is that both phases necessitate strong situational awareness, yet exhibit distinct priorities. In optimal conditions, agencies focus on establishing fundamental risk awareness, identifying early warning indicators, and strengthening systems. The grey sky, in contrast, highlights the real-time triage of data, rapid sense-making, and prompt resource allocation. These phases are interconnected: improved steady-state monitoring and preparedness facilitate more effective real-time adaptation during a crisis. By perceiving these stages as complimentary rather than isolated, stakeholders can transfer situational knowledge from blue sky to grey sky (e.g., identifying which neighborhoods are most susceptible to flooding or which electricity substations lack backup capacity). In contrast, lessons from grey sky occurrences reveal unforeseen vulnerabilities that contribute to enhancements during blue sky conditions.

\begin{table*}[!htbp]
\caption{\textbf{Focus and characteristics of blue-sky vs. gray-sky situational awareness (SA)}}
\label{table:blue-gray}
\setlength{\tabcolsep}{3pt}
\begin{tabularx}{\textwidth}{@{}|m{65pt}|m{150pt}|X|@{}}
\hline
Context & SA Focus and Characteristics & Examples of SA Information \\
\hline
Blue-Sky (Preparedness \& Mitigation) & Monitoring Normal Conditions and Emerging Risks
\begin{itemize}[leftmargin=*, nosep]
    \item Maintaining baseline awareness of resources and vulnerabilities.
    \item Emphasizing predictive and preventive information.
    \item Long-term situational understanding during "steady-state" operations.
\end{itemize}
 & Advanced Preparation for Disasters
 \begin{itemize}[leftmargin=*, nosep]
  \item Tracking weather forecasts and environmental indicators for early warning.
 \item Conducting training exercises and reviewing response plans.
 \item Monitoring infrastructure baseline status for anomalies detection.
 \item Awareness of community vulnerabilities and demographics.
 \item Identifying warning signs and mobilizing before disasters.
 \end{itemize}
 \\ \hline 
Gray-Sky (Event Onset \& Real-Time Crisis Response) & Real-Time Event Operations
\begin{itemize}[leftmargin=*, nosep]
    \item Achieving a COP.
    \item Emphasizing high-tempo data collection.
    \item Rapid dissemination and accuracy.
    \item SA often partial and dynamic.
    \item Requires continuous updates and triage of information.
\end{itemize}
 & Flood Operations Overview
 \begin{itemize}[leftmargin=*, nosep]
     \item Gathering real-time sensor readings or incident status reports from field.
     \item Monitoring casualties, trapped persons, and emergency services status.
     \item Providing immediate weather updates.
     \item Utilizing techniques like damage assessment teams, drone surveillance, and social media mining.
     \item Focusing on immediate actions and immediate needs.
 \end{itemize}
 \\ \hline 
\end{tabularx}
\end{table*}

\section{Roadmap for research and practice}
The preceding sections highlighted the significance of SA in both blue-sky and grey-sky situations, establishing it as a core resilience competency. SA enables rapid decision-making, fast information synthesis, and continuous risk assessment in both stable and emergency situations. It has been argued that if responders and communities fail to recognize, adapt to, and respond to evolving signals, then neither advanced hazard models nor comprehensive risk assessments will be adequate. Building on these premises, this section introduces a practical roadmap that integrates three interdependent dimensions: \textbf{technology}, \textbf{process}, and \textbf{people}.

\begin{itemize}
    \item \textbf{Technology}, such as, satellite imagery, Internet of Things (IoT) sensors, and AI-enhanced damage assessments offer multimodal data streams and analytical tools that facilitate real-time situational intelligence. Key priorities include integrating autonomous feeds (e.g., near-real-time hazard nowcasting) with interoperability standards that can make multiple agencies tap into a shared picture.
    \item \textbf{Processes} are the organizational procedures and practices that transform data into operational insights. Examples of structured sense-making techniques include micro-briefings, formalized data-sharing agreements, and clearly defined thresholds for raising alarms or deploying resources. Advanced technologies can overwhelm responders in the absence of reliable procedures, leading to fragmented awareness and disrupted coordination.
    \item \textbf{People} are ultimately responsible for interpreting data, validating anomalies, and adapting decisions in the face of uncertainty. SA prioritizes human factors. Human actors, including first responders, emergency operations center (EOC) personnel, legislators, and community volunteers, are responsible for data analysis, identifying irregularities, and adjusting processes in real-time. Training in distributed situational awareness, cross-agency collaboration, and cognitive load management ensures that procedural and technological innovations lead to efficient, context-aware actions.
\end{itemize}

\begin{figure}[!htbp]
\centering
\includegraphics[width=0.35\textwidth]{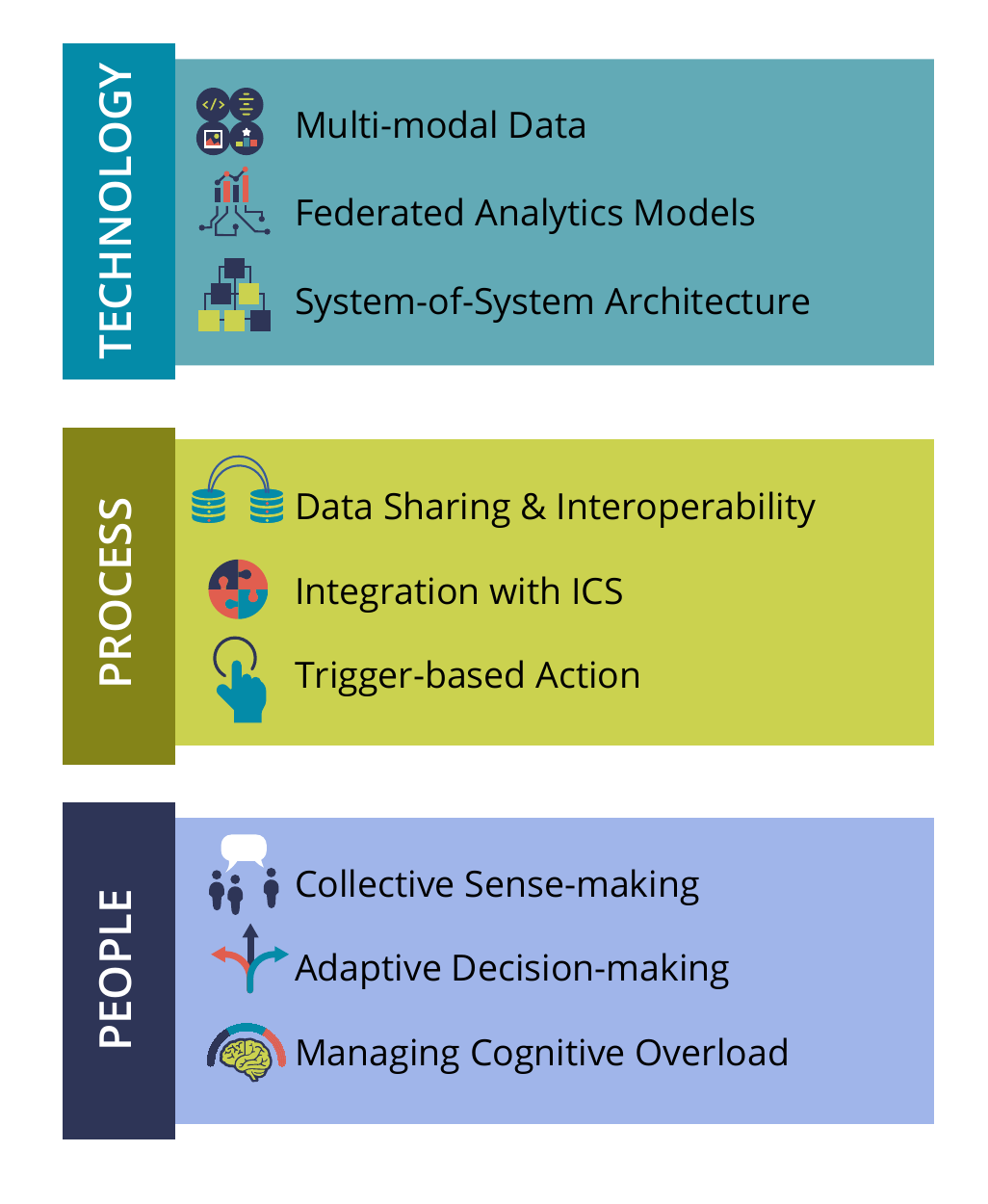}
\caption{A practical roadmap integrating technology, process, and people}\label{fig:tech-process-people}
\end{figure}

By uniting these three interdependent dimensions, the roadmap aims to operate SA aligned with both Endsley’s model (emphasizing perception, comprehension, and projection) and Distributed SA (stressing team and system perspectives). Section 4.1 highlights cutting‑edge technologies enabling situational intelligence; Section 4.2 delves into the processes for data integration and system interoperability; and Section 4.3 focuses on human factors critical to collective sense‑making. This integrated approach ensures that resilience practitioners and researchers can collaborate to close the gap between theoretical SA frameworks and the lived realities of disaster response, ultimately enhancing a community’s agility to confront the unanticipated.

\begin{figure*}[!htbp]
\centering
\includegraphics[width=0.9\textwidth]{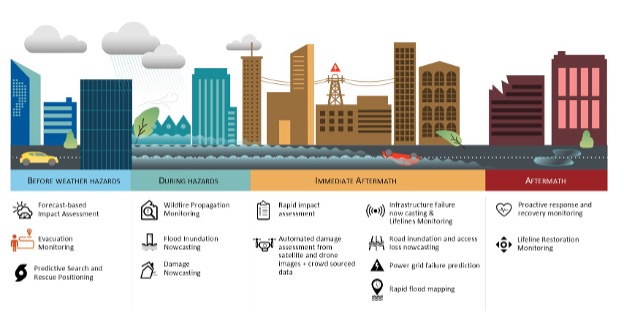}
\caption{Example SA capabilities across different hazard stages}\label{fig:SAexample}
\end{figure*}

\begin{figure*}[!htbp]
\centering
\includegraphics[width=0.75\textwidth]{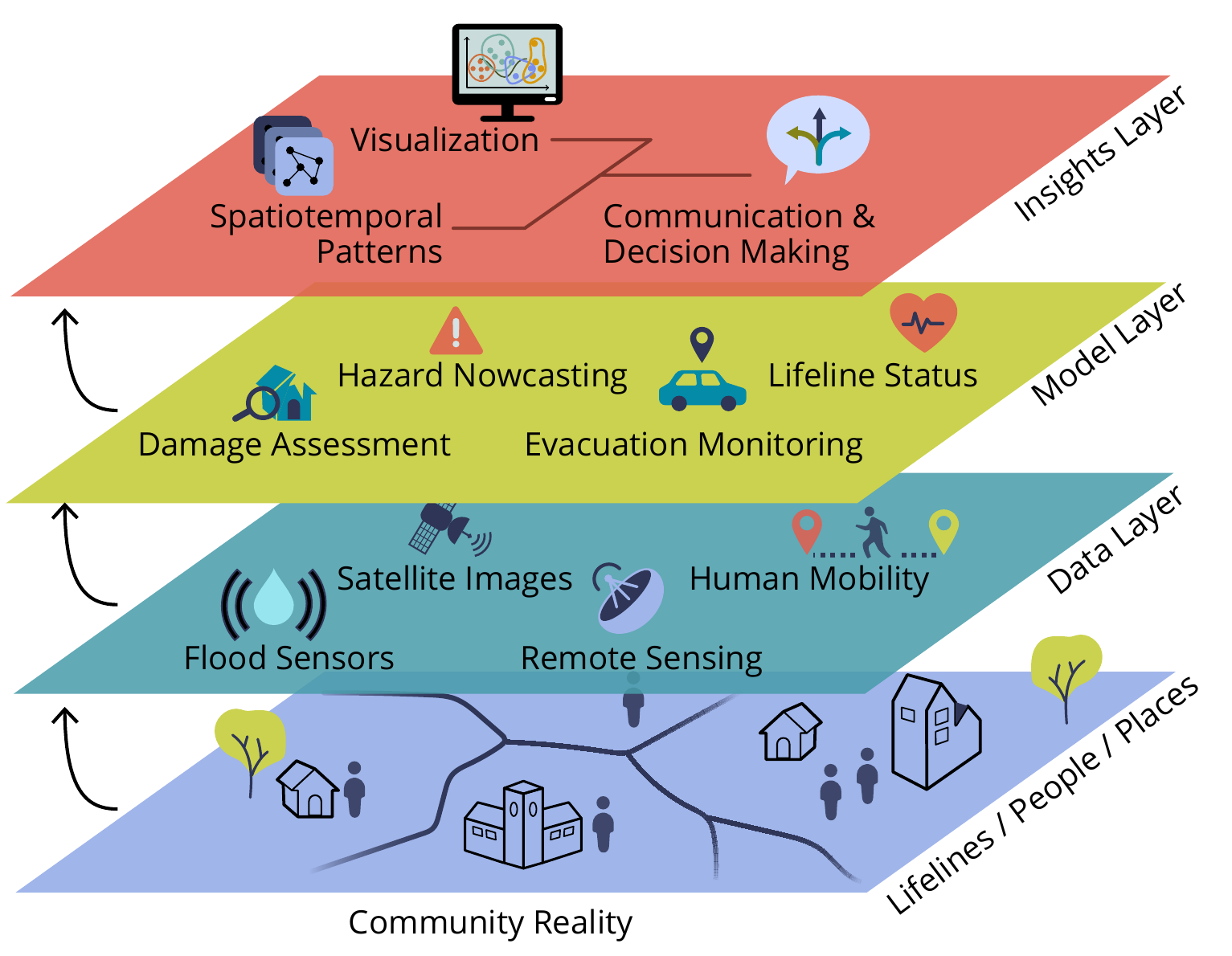}
\caption{Integrated insight framework for community disaster response}\label{fig:insight-model-data-community}
\end{figure*}

\subsection{Technologies: Harnessing multi modal data and AI across hazard stages}
In disaster situations, multi-modal data—which includes a combination of satellite imagery, sensor feeds, crowd reports, social media analytics, and traffic updates—is getting more and more important for providing dynamic, real-time responses to mitigate hazard risks and satisfy the community needs. This wide range of inputs can address the "blind spots" that appear when traditional risk models run into the complexity of unforeseen or compounding events. Agencies can extend SA beyond isolated or sporadic observations by integrating diverse datasets within a system-of-systems framework, as previously announced. This allows for continuous tackling of disaster escalation, frequently on an hourly or even minute-by-minute basis.

\subsubsection{Real-time hazard nowcasting}
Nowcasting systems, which are often updated at intervals shorter than daily or 12-hour forecasts, provide short-range projections about the current extent of hazards. For example, \citet{WIFIRE_Firemap_2025} combines fire detections, wind forecasts, and topography to predict wildfire spread in near real-time, while \citet{FloodMapp_NowCast_2025} updates flood inundation maps hourly by integrating live sensor and weather data. By sending dynamic hazard data straight to command centers, these solutions reduce "grey-sky" blind spots and enable quick reorientation in line with OODA Loop principles as circumstances change. These technologies can identify minor risk increases during "blue-sky" periods, prompting agencies to pre-deploy resources in anticipation of unforeseen storm intensification.

\subsubsection{Spatial computing for infrastructure interdependencies}
Modern societies depend on various interlinked "lifelines", such as electricity, water supply, telecommunications, and transportation systems. A single disruption can trigger cascading failures in critical services, cause collateral damage, and continuously exacerbate the overall response capability against crisis. By overlaying data on road closures, power grid sensor outputs, and demographics onto a single map, spatial computing tools like \citet{ResilitixAI_2025} draw attention to these interdependencies. By identifying important vulnerabilities, such as a substation supplying power to two large hospitals, agencies can concentrate on both proactive reinforcement, including the implementation of redundancy, and reactionary measures, such as deploying repair teams or generators. This methodology corresponds with DSA theory, wherein knowledge is distributed throughout diverse services, and only through their integration can responders completely grasp the extensive cascading implications.

\subsubsection{Automated damage \& impact assessment}
Immediately after a disruptive event, responders need swift and granular damage assessments to guide search and rescue operations. AI models, such as the SKAI tool \citep{WFP_SKAI_Project_2025} co-developed by the World Food Programme, analyze optical or radar satellite images to classify building damage within hours. Bridges at risk of collapsing can be automatically alerted by structural health monitoring sensors that are placed before a disaster strikes, and drones or IoT sensors can add close-up information to aerial views. This accelerates the triage during "grey-sky" events, allowing field teams to prioritize the most severely affected areas. Additionally, these technologies can be used for regular inspections or digital twin exercises during "blue-sky" periods, guaranteeing that maintenance or retrofitting takes place in advance of the next hazard.

\subsubsection{Near real-time monitoring of Evacuations and Population Movements}
Responders are uncertain about who is still in the hazard areas because evacuation compliance frequently deviates from official plans. Waze traffic \citep{Esri_Waze_CCP_2016} feeds offer real-time information on traffic congestion, while Facebook Disaster Maps \citep{Meta_Disaster_Maps_2025} and SafeGraph's location analytics \citep{safegraph2024} integrate phone data. Emergency managers and planners can proactively arrange additional shelters or reallocate resources by identifying neighborhoods that remain occupied despite evacuation orders, or by detecting unforeseen bottlenecks. These data streams can also identify regular commuter patterns and systemic vulnerabilities, like restricted egress routes, during "blue-sky" planning. These issues can then be resolved well before a crisis occurs.

\subsubsection{Predictive Search \& Rescue (SAR) Positioning}
Predictive impact models help authorities choose the best locations for rescue teams to deploy as hazard approaches. For instance, USGS PAGER \citep{earle2009prompt} forecasts casualties within minutes following an earthquake by correlating shaking intensity with known building fragilities, thereby prompting the dispatch of specialized urban SAR teams. FEMA's Hazus \citep{FEMA_Hazus_2025} can project structural damage from an impending storm and direct rapid water rescue units or medical supplies to potential hotspots. In Endsley's framework \citep{endsley1995measurement}, these pre-landfall intelligence tools are an example of Level 3 "projection"—the capacity to look beyond the immediate hazard to predict second-order effects. Over time, these predictions can inform "blue-sky" improvements, such as the implementation of stricter building codes or the strategic placement of SAR equipment.

\subsubsection{Integrated Lifeline Functionality Dashboards}
Disaster disruptions rarely remain confined to a single sector; a solitary event can simultaneously degrade power, water, roads, and communications. Multiple data feeds, such as hospital capacities, utility outage reports, and water main statuses, are combined into a color-coded overview by lifeline dashboards, like DHS's Community Lifeline Status System (CLSS) \citep{US2024HomelandSecurity}. By emphasizing critical or malfunctioning systems, these dashboards mitigate the "silo effect," bridging the theoretical understanding of baseline capacity with real-time information on outages or service restoration. Users, ranging from local EOCs to national coordination centers, can share a unified perspective, which facilitates consistent prioritization and resource allocation.

\subsubsection{Long Term Recovery Analytics}
After a disaster, recovery can take months or years, and equitable rebuilding requires keeping a SA of what has been restored and what is still damaged. Multi-modal data proves useful in these situations once more: credit card transactions or remote sensing identify persistent debris or business reopenings, while aggregated cellphone signals show population return rates. A blind spot that exists in many frameworks—the belief that recovery happens on its own after the "active" response phase—is addressed by this post-disaster visibility. By identifying lagging areas, officials can direct targeted support, ensuring that no group is permanently disadvantaged.

From "blue-sky" steady-state monitoring to "grey-sky" dynamic crisis response, these technological categories collectively demonstrate how real-time intelligence can inform every stage of a disaster. However, if agencies lack interoperable procedures or if overworked staff are unable to decipher the voluminous streams of social media signals, outage maps, and hazard forecasts, technology alone may prove ineffective. Therefore, creating a successful system of systems requires not just implementing these tools but also integrating them through cross-agency data agreements, open standards, and thorough training in distributed sense-making. This is the focus of the following sections, which examine how people and processes need to work together to make sure that these cutting-edge technologies actually improve SA rather than just making things more complicated.

\begin{table*}[!htbp]
\caption{Summary of Example SA Technologies in Disasters}
\label{table:summary SA}
\setlength{\tabcolsep}{3pt}
\begin{tabularx}{\textwidth}{@{}|m{55pt}|m{85pt}|X|m{55pt}|X|@{}}
\hline
Category & Example Technologies & Key Capabilities & Current Use Regions & Maturity (Research to Operational) \\ \hline
Hazard Nowcasting & 
\citet{FloodMapp_NowCast_2025,WIFIRE_Firemap_2025,ZestyAI_ZFIRE_2025}
& \begin{itemize}[leftmargin=*, nosep]
     \item Real-time hazard extent
     \item short-term prediction (flood depth, fire spread, etc.)
\end{itemize}
& North America, Australia, Europe & Mature \begin{itemize}[leftmargin=*, nosep]
     \item Operational for floods \& weather
     \item Wildfire models emerging (operational in some regions)
\end{itemize} \\ \hline 
Infrastructure Functional Failure & 
\citet{ResilitixAI_2025,HHS_emPOWER_program,OneConcern_ResiliencePlatform_2025}
 & \begin{itemize}[leftmargin=*, nosep]
     \item Mapping critical nodes and system-of-systems behavior
     \item Identifying crucial facilities and predicting cascading failures
\end{itemize}
 & USA, Japan (pilot), global interest & Mixed \begin{itemize}[leftmargin=*, nosep]
     \item emPOWER fully operational
     \item others in pilot/early adoption
\end{itemize} \\ \hline 
Rapid Damage Assessment & \citet{WFP_SKAI_Project_2025,Mapfre_DronesSatellites_2025}; IoT sensors
 & \begin{itemize}[leftmargin=*, nosep]
     \item Automated identification of damage and outages within hours
     \item Multi-source (imagery, sensors) for comprehensive impact mapping
\end{itemize}
& Global (major disaster responses by UN, FEMA, etc.) & Reaching Operational \begin{itemize}[leftmargin=*, nosep]
     \item Satellite/drone AI used in recent disasters with high accuracy
     \item IoT sensors usage growing
\end{itemize} \\ \hline 
Evacuation \& Population Monitoring & \citet{Meta_Disaster_Maps_2025,ResilitixAI_2025,safegraph2024,Esri_Waze_CCP_2016}
 & \begin{itemize}[leftmargin=*, nosep]
     \item Near-real-time tracking of population movements, evacuation compliance, and sheltering via digital footprints
\end{itemize} & Global (Facebook); US/EU (Waze) & Operational \begin{itemize}[leftmargin=*, nosep]
     \item Being integrated into emergency dashboards and protocols in many places
\end{itemize} \\ \hline
Predictive SAR Deployment & \citet{earle2009prompt,FEMA_CommunityLifelines_2024,yan2025mobirescue}
& \begin{itemize}[leftmargin=*, nosep]
     \item Forecast likely hardest-hit areas and optimal resource placement
     \item Enhances speed and effectiveness of SAR and relief
\end{itemize}
& Global (PAGER); US (Hazus); pilot (AI dispatch) & \begin{itemize}[leftmargin=*, nosep]
     \item Operational for impact modeling
     \item Research phase for AI dispatch
\end{itemize} \\ \hline
Lifeline Status Monitoring & \citet{ResilitixAI_2025,cusec_clss_2025}; Utility outage systems; Satellite outages & \begin{itemize}[leftmargin=*, nosep]
     \item Live status updates of critical services (power, water, transport) in disaster area
     \item Supports prioritized restoration and mutual aid
\end{itemize}
& US (CLSS, DOE); Global (satellite) & Operational \begin{itemize}[leftmargin=*, nosep]
     \item many parts in use, with integration platforms (CLSS) just now launching to unify them
\end{itemize} \\ \hline
Recovery Tracking & Mobility return analysis; Nighttime lights; Economic data trackers
& \begin{itemize}[leftmargin=*, nosep]
     \item Ongoing monitoring of community recovery: population, business, infrastructure functionality over months
     \item Flags areas needing continued support
\end{itemize} & US (several studies); global (satellite \& pilot projects) & Early Operational \begin{itemize}[leftmargin=*, nosep]
     \item Used in assessments and decision-making by some agencies
     \item But not yet universal
\end{itemize} \\ \hline
\end{tabularx}
\end{table*}

\subsubsection{Integrating Multiple Technologies via a System of Systems Framework for Multi Agency SA}
Individual SA tools, such as flood nowcasters, AI-based damage assessments, and infrastructure dashboards, offer distinct capabilities. However, effective SA across multiple agencies is most often achieved when these tools are integrated within a broader system-of-systems framework. Rather than requiring all stakeholders to adopt a single, uniform platform, this approach employs a central orchestration layer that aggregates data and analytics from multiple, independently operated sources. Each of these sources retains its own technical methodology and governance structure. By emphasizing open standards and distributed data flows, agencies and utilities can preserve their established workflows while concurrently contributing to and benefiting from a shared, real-time operational picture during crisis events.

Federated data ownership is a key part of the system-of-systems approach. In this model, each participant, whether it is a public agency, a private utility, or a nongovernmental organization, has full control over its own data. Because of this, no one organization sets data policies or enforces the same standards. Instead, local experts can update or fix their information, which makes the data more accurate. This federated architecture works especially well for events that involve more than one jurisdiction, where command structures are often changing and adapting. Interoperable interfaces also make it easy to add hazard feeds, infrastructure datasets, and population rasters to shared visual environments without any problems. These interfaces often adhere to widely accepted protocols such as OGC WMS/WFS or RESTful APIs. Dynamic service-to-service connections ensure that updates from source systems are reflected almost instantaneously, eliminating reliance on static data transfers and enabling a continuously updated view of the unfolding situation. Modularity stands as an equally critical design principle, permitting advanced analytic components—such as predictive outage models, seismic event calculators, or public health monitoring tools—to be incorporated without necessitating extensive reconfiguration of the core infrastructure. Each analytic module operates as an independent service, publishing its outputs to the orchestration layer in real time. This architecture not only supports the integration of emerging AI-driven tools but also facilitates the replacement of obsolete components with minimal system disruption.

Using open data makes this environment even better by encouraging non-proprietary data feeds, which lowers legal and security risks. Sources that are open to the public help get around common licensing and privacy issues, making the emergency coordination ecosystem more open and flexible. The Energy Awareness and Resiliency Standardized Services (EARSS) platform \cite{ORNL_EARSS_Operational}, which was made at Oak Ridge National Laboratory, is an example of how these ideas can be used in the energy and critical infrastructure sectors. EARSS combines a wide range of feeds into a single geodatabase. These feeds include utility outage dashboards, publicly available hazard data like NOAA hurricane tracks and USGS ShakeMaps, demographic rasters, and infrastructure registries. One important thing about EARSS is that it uses data that is available to the public, which makes it easier for agencies to share data during emergencies. Agencies can access these live map layers via standardized OGC services or through a dedicated Google Earth "Connector." For example, EARSS provides near-real-time spatial overlays that local EOCs, state fusion centers, and federal coordination hubs can use to make sure that everyone is on the same page about how to use resources and communicate with the public.

Beyond simply distributing timely hazard and impact information, the system-of-systems model enhances multi-agency coordination in three principal ways. First, it enables agencies to participate with minimal integration overhead by allowing them to contribute datasets—such as local substation geospatial data—without forfeiting data ownership or modifying existing IT configurations. Second, responders who are already using established SA platforms like WebEOC or ArcGIS dashboards will be able to access new data in familiar operational environments thanks to standardized interfaces. Third, and most importantly, real-time outputs like estimated power outages and projected restoration timelines help you make decisions about how to run your business right away. This makes it easier to put together different crisis snapshots on your own. Third, and most importantly, real-time outputs, such as estimated power outages and projected restoration timelines, help with operational decision-making directly, which cuts down on the time it takes to put together different crisis snapshots on your own. The system-of-systems paradigm can be used in other areas as well, such as flood forecasting, AI-based assessments of structural damage, disease surveillance, and community-sourced social media analytics. EARSS shows how this works in the energy sector. Agencies work together to build a constantly improving COP by putting together different models into a single, evolving framework. Each new piece of information improves overall SA, since contributors can still control their own data and send verified updates in real time without giving up control of their data.

Ultimately, federated data governance, adherence to open standards, and a modular analytics framework collectively establish a scalable, flexible, and legally unobstructed environment for inter-agency collaboration. This system-of-systems architecture holds the potential to deliver the high-resolution, near-real-time intelligence necessary to manage the complex uncertainties and cascading failures characteristic of major disasters.

\subsection{Processes: Strengthening Workflows for Technology-Enabled SA}
Even the most advanced SA technologies won't work well if they aren't built upon efficient workflows and protocols that make sure they deliver timely, useful intelligence instead of just a lot of data or isolated silos. As disasters today become more complicated in terms of space and time and affect more than one jurisdiction, it is very important for engineers to come up with systematic ways to combine data, help with decisions, and coordinate between agencies. Standardized protocols are needed for modern disaster scenarios that can turn different types of data streams from many sources into useful information for coordinated emergency response operations. The next few sections show a complete framework for how well-designed information systems can turn raw sensor data and observational inputs into synchronized response protocols that improve safety and reduce damage to infrastructure.

\subsubsection{Centralized vs. Distributed Data Hubs}
Determining the best data aggregation environment, whether SA data should be dispersed among various agency platforms or consolidated within a centralized information infrastructure, is the main architectural consideration. A centralized data portal architecture enables the integration of heterogeneous real-time data streams, including hydrodynamic flood nowcasts, multi-spectral satellite imagery, and social media sentiment analytics, into a unified geospatial visualization interface. This makes sure that all of the dashboards are the same. Some agencies may want to keep their own analytical tools and only share results that have been checked. Organizations keep their operational independence while making sure that everyone involved understands the issue fully by clearly stating which data is kept locally and which is managed centrally. This includes things like common geospatial layers and near-real-time hazard feeds. This way of designing things also helps with information overload because it lets centralized dashboards show only anomalies or threshold exceedances instead of every raw sensor reading.

\subsubsection{Sense Making Protocols and Scheduled Briefings}
Having a lot of data doesn't automatically help you understand what's going on; making sense of it is still a human-driven, ongoing process. Regularly scheduled "micro-briefings" or "huddles" facilitate the transformation of raw data streams into concise situational summaries, particularly in high-pressure scenarios. An EOC may provide updates every hour, with each update lasting ten minutes. During these updates, leaders from operations, logistics, and planning reconcile discrepancies in reports, reassess resource utilization, and revise the COP as necessary. Adding these analytical protocols to incident management workflows makes it easier to quickly find anomalies and raise the alarm about important data changes, like sudden rises in water levels or spikes in demand for emergency services. These shorter decision cycles make it easier for operators to think by spreading out the processing of information, which also lowers the chance of missing important signals.

\subsubsection{Integrating SA with the Incident Command System (ICS)}
The Incident Command System (ICS) or similar command frameworks are used by many places to coordinate disaster response by more than one agency. Hazard nowcasting can help the Planning Section's intelligence unit directly, and near real-time damage assessments can help the Operations Section make decisions about how to use its resources. To prevent the creation of isolated "technology silos," it is essential to ensure that SA technologies integrate with ICS roles and processes. A designated Liaison Officer may oversee data requests from external partners to clarify data ownership and management. By correlating technological inputs with defined ICS responsibilities, the information flow may be tailored for each role, preventing overload and ensuring timely updates for all pertinent parties. Individuals at the EOC are more inclined to trust and utilize these solutions if they align seamlessly with existing practices.

\subsubsection{Interoperability and Data Sharing Agreements}
Without clear rules for fast and easy data exchange, it's hard to fully use technological capabilities. Before a disaster, Memoranda of Understanding (MOUs) or mutual aid agreements usually set the main rules, such as how often data should be updated, what formats are best (like GeoJSON or OGC services), and what rules there are for governance (like data retention policies or classification levels). These agreements help keep negotiations to a minimum and stop the creation of "island" systems that can't talk to each other. When data is combined into a single operating picture, it is easier to understand because everyone uses the same names and metrics, like the same road closure statuses or the same flood depth units. Interoperability helps everyone understand what's going on by getting rid of agency-specific jargon and unnecessary categories. This is important because quick decisions can save lives.

\subsubsection{Rapid Escalation and Trigger Based Actions}
Time is usually the most important thing when disasters happen quickly. When certain things happen, like water levels going above a certain point or an AI model predicting that structural damage will be more than 30\% in an area, these things can automatically set off escalations in incident response. This could mean sending in specialized response teams or raising the EOC's activation level. These triggers go off right away when the data is perfectly integrated. People don't have to figure them out or follow orders that are in a hierarchy. This makes sure that a lot of people in charge do things quickly and in a planned way. So, well-defined processes make it easy to go from being aware of a threat to taking quick action to close the gap between seeing a threat and doing something about it.

\subsubsection{Regular Drills and Training for Data Integration}
The EOC, field response teams, or partner non-governmental organizations don't know how to use advanced analytical interfaces well enough, which leads to poor system performance and underused technological capabilities. Comprehensive joint training protocols must include requirements for processing multi-modal data streams, where participants show they can interpret data in near-real-time, run predictive models, and create standardized situation reports in a controlled crisis simulation setting. These practice sessions teach more than just how to click buttons. They also point out problems with the user interface, point out possible problems with data flow, and get teams to work together to sort and rate data. These exercises test interoperability by including outside partners, such as private utilities and volunteer groups. This makes sure that everyone can share data without any problems. Because data-driven exercises that promote a culture of analytical proficiency are always being used, better SA technologies are now a necessary part of crisis workflows instead of an afterthought.

\subsubsection{Building Continuous Improvement Loops}
After each disaster, a systematic review is done to find problems with SA, such as escalation triggers that don't match up, reports of outages that don't match up, or delays in getting the word out about flooded roads. Continuous improvement is possible when you write down what you learned and change your data-sharing agreements or Standard Operating Procedures (SOPs) as needed. For example, a city might discover that they didn't use drone images enough or that they set resource pre-staging thresholds too low. Implementing these lessons in future protocols, such as clarifying escalation criteria, standardizing data definitions, and conducting more frequent "micro-briefings" among agencies, enables businesses to adhere to the principle of iterative learning, which asserts that all individuals should have access to the same information. Over time, this reflective approach ensures that each incident not only evaluates but also enhances the system's capacity to provide real-time SA.

\subsection{People: Closing the Loop with Human Judgment and Collective Cognition}
The operational effectiveness of advanced SA technologies and optimized workflow architectures ultimately depends on how well people can understand data, resolve conflicting information, and make quick decisions that are relevant to the situation. Human operators function as critical system components that transform heterogeneous data inputs into executable response protocols during disaster scenarios characterized by information saturation, dynamic operational requirements, and emergent hazard conditions that exceed standard response plan parameters. Through systematic alignment of technological systems with established principles of cognitive load management, distributed cognition frameworks, and adaptive decision-making protocols, emergency management agencies can achieve optimal performance from DSA implementations.

\subsubsection{Managing Cognitive Overload under Crisis Conditions}
When trained workers are under a lot of stress, they may not be able to think as clearly as possible, especially with too much information from a lot of different sources, like distributed sensor networks, crowdsourced alert systems, and machine learning-based predictive models. Automated detection algorithms can find statistical outliers, such as when water levels suddenly rise or when the electrical grid fails in a chain reaction. But people still need to check alerts and pick the best response protocol. Operators can see a lot of information, but they can't use it to make decisions when there is too much of it. This happens when important signals get lost in alert streams that happen a lot but aren't very important. To keep operators' performance at an acceptable level, companies need to set up systematic cognitive load management protocols. Setting up structured analytical rest breaks, using threshold-based filtering algorithms that only show parameter exceedances above set criteria, and using rotational assignment protocols for functions that need close monitoring are all proven ways to lower risk. Endsley's well-known three-level SA framework for running complex systems says that keeping operators mentally alert makes it easier for them to see, understand, and predict things.

\subsubsection{Collective Sense Making and Team Coordination}
No one role, department, or technological data stream can hold all the relevant information during a disaster, so everyone needs to think together. When field reports, GIS overlays, and public health intelligence come together to create a shared COP, which is often made possible by quick conversation and cross-verification, SA becomes effective. Some team rules that help everyone stay on the same page when new information comes in are "micro-briefings" or "huddles." This makes sure that everyone is always updating their mental picture of the crisis in the same way as everyone else. This is how DSA works when people work together. There isn't one central command center in DSA where information and knowledge are kept. Instead, they are spread out among many people and information artifacts. It's important to have open lines of communication so that employees can question AI-generated outputs or assumptions that don't match what's really going on. This helps lower the risk of automation bias and dependence on a single data source.

\subsubsection{Encouraging Improvisation and Adaptive Decision Making}
Standardized answers come from carefully planned procedures, but you need to be able to think on your feet when things don't go as planned. More than one threat can hit a community at the same time, like a hurricane followed by a series of infrastructure failures. They can also find themselves in situations they didn't expect. Frontline responders must be ready to break the rules if real-time SA shows that a different course of action is more urgent or effective. When companies promote decentralized decision-making, local teams can quickly change their plans without having to wait for permission from higher-ups. When responders use both AI data, like predictions of new flooded evacuation routes, and local knowledge, they can make better backup plans or move resources around more efficiently. When disasters happen, there are a lot of things that are unknown. To deal with these, you need to find a balance between letting people make their own choices and having structured operational frameworks.

\subsubsection{Building Trust and Cross Agency Collaboration}
During multi-agency disaster responses, such as floods, wildfires, and pandemics, local governments, non-governmental organizations, private utilities, and volunteer groups often work together. For SA to work, these different partners need to trust each other. If any stakeholder's data is met with doubt or if cultural or legal barriers make it hard to share information, the overall picture of the situation stays incomplete. Formal agreements to share data set up a structure, but trust is mostly built through regular cooperative exercises and interactions that show that each person can be trusted and knows a lot about their field. The presence of liaison officers, embedded personnel, or rotating task forces further reinforces organizational cohesion within a shared sense-making process. Consequently, the successful technical integration of hazard feeds and analytics is more likely to yield a cohesive and widely accepted COP when strong interpersonal relationships underpin the collaboration.

\subsubsection{Training and Institutional Learning}
Everyone, from EOC analysts to field responders to community liaisons, needs to keep getting training so that they can read real-time data, use digital platforms, and properly escalate triggers during a disaster. Simulated drills that mimic different data situations, like a rise in social media activity or sensor problems, help find problems with how teams deal with conflicting reports and keep everyone updated when things get tough. It is just as important to learn institutional knowledge from real-life events. To help training programs get better, after-action reviews should keep track of both successes, like avoiding crises by quickly adapting, and failures, like information overload or missing anomalies. Over time, this iterative feedback process helps build a workforce that can use multi-modal intelligence to make timely, well-informed operational decisions.

\section{Concluding Remarks}
SA is an important part of modern disaster resilience frameworks. It helps to fill in operational gaps that happen when hazard evolution rates are faster than standard modeling and planning response capabilities. This study has shown that SA is a key engineering principle in resilience system architectures. It has shown that it can help fix operational blind spots that happen when hazard evolution rates are faster than standard modeling and planning response capabilities.

The main point of this work is to show how to systematically connect established SA theoretical frameworks, like Endsley's three-level cognitive model and DSA architectures, with real-world problems in multi-agency coordination systems, dealing with too much information, and responding to new threats. This analysis shows how hydrodynamic nowcasting systems, social media analytics platforms, and critical infrastructure monitoring dashboards can work together by using standardized data exchange protocols. It does this by combining the cognitive processes of perception, comprehension, and projection within system-of-systems engineering frameworks. To use these technologies, they need to be integrated in a planned way with strong operational processes that set up data-sharing frameworks, structured micro-briefing protocols, and make sure that automated alert systems start the right escalation procedures. The ability of an organization to effectively use advanced intelligence systems is based on human factors engineering principles such as cognitive load management, distributed team sense-making protocols, and adaptive decision-making frameworks. Technology, process, and people all work together as parts of a SA architecture to improve system performance and create coherent operational intelligence in tough situations.

Future research directions include creating quantitative performance metrics that measure the effectiveness and rates of organizational distributed intelligence fusion, creating interpretable interfaces and explainable algorithm architectures for AI systems, and improving public engagement methods to make SA systems work better in communities that are often left out or are at high risk. SA systems give you the operational flexibility and analytical clarity you need to stop damage from spreading in an operational environment where hazards are becoming more common and failures can happen in a chain reaction.

This study calls for a major change in the way disaster resilience engineering is done. It says that we should move away from narrow approaches that focus on mitigation and toward broad SA-centric frameworks that are used in all phases of disaster management, from constant monitoring during normal operations to real-time adaptive response in critical situations and proactive recovery implementation. SA systems need to be constantly improved, not just as technological improvements, but also as unifying engineering principles that systematically bring together technology, process, and human factors to support resilient community infrastructure systems around the world.





\printcredits

\bibliographystyle{cas-model2-names}

\bibliography{cas-refs}

\begin{thebibliography}{55}
\expandafter\ifx\csname natexlab\endcsname\relax\def\natexlab#1{#1}\fi
\providecommand{\url}[1]{\texttt{#1}}
\providecommand{\href}[2]{#2}
\providecommand{\path}[1]{#1}
\providecommand{\DOIprefix}{doi:}
\providecommand{\ArXivprefix}{arXiv:}
\providecommand{\URLprefix}{URL: }
\providecommand{\Pubmedprefix}{pmid:}
\providecommand{\doi}[1]{\href{http://dx.doi.org/#1}{\path{#1}}}
\providecommand{\Pubmed}[1]{\href{pmid:#1}{\path{#1}}}
\providecommand{\bibinfo}[2]{#2}
\ifx\xfnm\relax \def\xfnm[#1]{\unskip,\space#1}\fi
\bibitem[{Brown et~al.(2008)Brown, Son, Aziz and Pe{\~n}a-Mora}]{brown2008supporting}
\bibinfo{author}{Brown, S.}, \bibinfo{author}{Son, J.}, \bibinfo{author}{Aziz, Z.}, \bibinfo{author}{Pe{\~n}a-Mora, F.}, \bibinfo{year}{2008}.
\newblock \bibinfo{title}{Supporting disaster response and recovery through improved situation awareness}.
\newblock \bibinfo{journal}{Structural survey} \bibinfo{volume}{26}, \bibinfo{pages}{411--425}.
\bibitem[{Chen et~al.(2008)Chen, Sharman, Rao and Upadhyaya}]{chen2008coordination}
\bibinfo{author}{Chen, R.}, \bibinfo{author}{Sharman, R.}, \bibinfo{author}{Rao, H.R.}, \bibinfo{author}{Upadhyaya, S.J.}, \bibinfo{year}{2008}.
\newblock \bibinfo{title}{Coordination in emergency response management}.
\newblock \bibinfo{journal}{Communications of the ACM} \bibinfo{volume}{51}, \bibinfo{pages}{66--73}.
\bibitem[{Cheng et~al.(2023)Cheng, Khajwal, Behzadan and Noshadravan}]{cheng2023probabilistic}
\bibinfo{author}{Cheng, C.S.}, \bibinfo{author}{Khajwal, A.B.}, \bibinfo{author}{Behzadan, A.H.}, \bibinfo{author}{Noshadravan, A.}, \bibinfo{year}{2023}.
\newblock \bibinfo{title}{A probabilistic crowd--ai framework for reducing uncertainty in postdisaster building damage assessment}.
\newblock \bibinfo{journal}{Journal of Engineering Mechanics} \bibinfo{volume}{149}, \bibinfo{pages}{04023059}.
\bibitem[{Coleman et~al.(2023)Coleman, Liu, Zhao and Mostafavi}]{coleman2023lifestyle}
\bibinfo{author}{Coleman, N.}, \bibinfo{author}{Liu, C.}, \bibinfo{author}{Zhao, Y.}, \bibinfo{author}{Mostafavi, A.}, \bibinfo{year}{2023}.
\newblock \bibinfo{title}{Lifestyle pattern analysis unveils recovery trajectories of communities impacted by disasters}.
\newblock \bibinfo{journal}{Humanities and Social Sciences Communications} \bibinfo{volume}{10}, \bibinfo{pages}{1--13}.
\bibitem[{Curnin et~al.(2015)Curnin, Owen, Paton, Trist and Parsons}]{curnin2015role}
\bibinfo{author}{Curnin, S.}, \bibinfo{author}{Owen, C.}, \bibinfo{author}{Paton, D.}, \bibinfo{author}{Trist, C.}, \bibinfo{author}{Parsons, D.}, \bibinfo{year}{2015}.
\newblock \bibinfo{title}{Role clarity, swift trust and multi-agency coordination}.
\newblock \bibinfo{journal}{Journal of Contingencies and Crisis Management} \bibinfo{volume}{23}, \bibinfo{pages}{29--35}.
\bibitem[{Earle et~al.(2009)Earle, Wald, Jaiswal, Allen, Hearne, Marano, Hotovec and Fee}]{earle2009prompt}
\bibinfo{author}{Earle, P.S.}, \bibinfo{author}{Wald, D.J.}, \bibinfo{author}{Jaiswal, K.S.}, \bibinfo{author}{Allen, T.I.}, \bibinfo{author}{Hearne, M.G.}, \bibinfo{author}{Marano, K.D.}, \bibinfo{author}{Hotovec, A.J.}, \bibinfo{author}{Fee, J.M.}, \bibinfo{year}{2009}.
\newblock \bibinfo{title}{Prompt assessment of global earthquakes for response (pager): A system for rapidly determining the impact of earthquakes worldwide}.
\newblock \bibinfo{journal}{US Geological Survey Open-File Report} \bibinfo{volume}{1131}, \bibinfo{pages}{15}.
\bibitem[{Endsley(1995)}]{endsley1995measurement}
\bibinfo{author}{Endsley, M.R.}, \bibinfo{year}{1995}.
\newblock \bibinfo{title}{Measurement of situation awareness in dynamic systems}.
\newblock \bibinfo{journal}{Human factors} \bibinfo{volume}{37}, \bibinfo{pages}{65--84}.
\bibitem[{Endsley(2017)}]{endsley2017toward}
\bibinfo{author}{Endsley, M.R.}, \bibinfo{year}{2017}.
\newblock \bibinfo{title}{Toward a theory of situation awareness in dynamic systems}, in: \bibinfo{booktitle}{Situational awareness}. \bibinfo{publisher}{Routledge}, pp. \bibinfo{pages}{9--42}.
\bibitem[{Esparza et~al.(2025)Esparza, Li, Ma and Mostafavi}]{esparza2025ai}
\bibinfo{author}{Esparza, M.}, \bibinfo{author}{Li, B.}, \bibinfo{author}{Ma, J.}, \bibinfo{author}{Mostafavi, A.}, \bibinfo{year}{2025}.
\newblock \bibinfo{title}{Ai meets natural hazard risk: A nationwide vulnerability assessment of data centers to natural hazards and power outages}.
\newblock \bibinfo{journal}{International Journal of Disaster Risk Reduction} , \bibinfo{pages}{105583}.
\bibitem[{{Esri}(2016)}]{Esri_Waze_CCP_2016}
\bibinfo{author}{{Esri}}, \bibinfo{year}{2016}.
\newblock \bibinfo{title}{Esri Partners with Waze Connected Citizens Program to Deliver Unprecedented Open Data‑Sharing Options to Governments}.
\newblock \bibinfo{type}{Press Release} \bibinfo{number}{—}. Esri. \bibinfo{address}{Redlands, CA}.
\newblock \URLprefix \url{https://www.esri.com/about/newsroom/announcements/esri-waze-connected-citizens-program-unprecedented-open-datasharing-governments}.
\bibitem[{Fan et~al.(2020a)Fan, Jiang and Mostafavi}]{fan2020social}
\bibinfo{author}{Fan, C.}, \bibinfo{author}{Jiang, Y.}, \bibinfo{author}{Mostafavi, A.}, \bibinfo{year}{2020}a.
\newblock \bibinfo{title}{Social sensing in disaster city digital twin: Integrated textual--visual--geo framework for situational awareness during built environment disruptions}.
\newblock \bibinfo{journal}{Journal of Management in Engineering} \bibinfo{volume}{36}, \bibinfo{pages}{04020002}.
\bibitem[{Fan et~al.(2020b)Fan, Wu and Mostafavi}]{fan2020hybrid}
\bibinfo{author}{Fan, C.}, \bibinfo{author}{Wu, F.}, \bibinfo{author}{Mostafavi, A.}, \bibinfo{year}{2020}b.
\newblock \bibinfo{title}{A hybrid machine learning pipeline for automated mapping of events and locations from social media in disasters}.
\newblock \bibinfo{journal}{IEEE Access} \bibinfo{volume}{8}, \bibinfo{pages}{10478--10490}.
\bibitem[{Fan et~al.(2021)Fan, Zhang, Yahja and Mostafavi}]{fan2021disaster}
\bibinfo{author}{Fan, C.}, \bibinfo{author}{Zhang, C.}, \bibinfo{author}{Yahja, A.}, \bibinfo{author}{Mostafavi, A.}, \bibinfo{year}{2021}.
\newblock \bibinfo{title}{Disaster city digital twin: A vision for integrating artificial and human intelligence for disaster management}.
\newblock \bibinfo{journal}{International journal of information management} \bibinfo{volume}{56}, \bibinfo{pages}{102049}.
\bibitem[{Farahmand et~al.(2022)Farahmand, Liu, Dong, Mostafavi and Gao}]{farahmand2022network}
\bibinfo{author}{Farahmand, H.}, \bibinfo{author}{Liu, X.}, \bibinfo{author}{Dong, S.}, \bibinfo{author}{Mostafavi, A.}, \bibinfo{author}{Gao, J.}, \bibinfo{year}{2022}.
\newblock \bibinfo{title}{A network observability framework for sensor placement in flood control networks to improve flood situational awareness and risk management}.
\newblock \bibinfo{journal}{Reliability Engineering \& System Safety} \bibinfo{volume}{221}, \bibinfo{pages}{108366}.
\bibitem[{{Federal Emergency Management Agency (FEMA)}(2025a)}]{FEMA_Hazus_2025}
\bibinfo{author}{{Federal Emergency Management Agency (FEMA)}}, \bibinfo{year}{2025}a.
\newblock \bibinfo{title}{Hazus: U.s. standardized risk‑assessment tool for natural hazards}.
\newblock \bibinfo{howpublished}{Online}.
\newblock \URLprefix \url{https://www.fema.gov/flood-maps/products-tools/hazus}.
\bibitem[{{Federal Emergency Management Agency (FEMA)}(2025b)}]{FEMA_PowerOutage_2025}
\bibinfo{author}{{Federal Emergency Management Agency (FEMA)}}, \bibinfo{year}{2025}b.
\newblock \bibinfo{title}{Power Outage Incident Annex to the Response and Recovery Federal Interagency Operational Plans: Managing the Cascading Impacts from a Long‑Term Power Outage}.
\newblock \bibinfo{type}{Technical Annex}. Federal Emergency Management Agency. \bibinfo{address}{Washington, DC}.
\newblock \URLprefix \url{https://www.fema.gov/sites/default/files/documents/fema_incident-annex_power-outage.pdf}.
\bibitem[{{FEMA}(2024)}]{FEMA_CommunityLifelines_2024}
\bibinfo{author}{{FEMA}}, \bibinfo{year}{2024}.
\newblock \bibinfo{title}{Community lifelines: 2024 doctrine update}.
\newblock \bibinfo{howpublished}{FEMA official website}.
\newblock \URLprefix \url{https://www.fema.gov/emergency-managers/practitioners/lifelines}. \bibinfo{note}{document detailing the 2024 update to the Community Lifelines construct, launched by FEMA in March 2024; includes updated doctrine for response and recovery across the whole community}.
\bibitem[{{FloodMapp}(2025)}]{FloodMapp_NowCast_2025}
\bibinfo{author}{{FloodMapp}}, \bibinfo{year}{2025}.
\newblock \bibinfo{title}{Floodmapp nowcast: Real‑time flood inundation mapping service}.
\newblock \bibinfo{howpublished}{Online}.
\newblock \URLprefix \url{https://www.floodmapp.com/nowcast}.
\bibitem[{Freitas et~al.(2020)Freitas, Borges and Carvalho}]{freitas2020conceptual}
\bibinfo{author}{Freitas, D.P.}, \bibinfo{author}{Borges, M.R.}, \bibinfo{author}{Carvalho, P.V.R.d.}, \bibinfo{year}{2020}.
\newblock \bibinfo{title}{A conceptual framework for developing solutions that organise social media information for emergency response teams}.
\newblock \bibinfo{journal}{Behaviour \& Information Technology} \bibinfo{volume}{39}, \bibinfo{pages}{360--378}.
\bibitem[{{G\&H International}(2025)}]{cusec_clss_2025}
\bibinfo{author}{{G\&H International}}, \bibinfo{year}{2025}.
\newblock \bibinfo{title}{Community lifeline status system (clss)}.
\newblock \bibinfo{howpublished}{ArcGIS Hub website}.
\newblock \URLprefix \url{https://clss-cusec.hub.arcgis.com/}. \bibinfo{note}{accessed: 2025‑07‑31}.
\bibitem[{Harrald and Jefferson(2007)}]{harrald2007shared}
\bibinfo{author}{Harrald, J.}, \bibinfo{author}{Jefferson, T.}, \bibinfo{year}{2007}.
\newblock \bibinfo{title}{Shared situational awareness in emergency management mitigation and response}, in: \bibinfo{booktitle}{2007 40th Annual Hawaii International Conference on System Sciences (HICSS'07)}, \bibinfo{organization}{IEEE}. pp. \bibinfo{pages}{23--23}.
\bibitem[{Hernantes et~al.(2013)Hernantes, Rich, Laug{\'e}, Labaka and Sarriegi}]{hernantes2013learning}
\bibinfo{author}{Hernantes, J.}, \bibinfo{author}{Rich, E.}, \bibinfo{author}{Laug{\'e}, A.}, \bibinfo{author}{Labaka, L.}, \bibinfo{author}{Sarriegi, J.M.}, \bibinfo{year}{2013}.
\newblock \bibinfo{title}{Learning before the storm: Modeling multiple stakeholder activities in support of crisis management, a practical case}.
\newblock \bibinfo{journal}{Technological Forecasting and Social Change} \bibinfo{volume}{80}, \bibinfo{pages}{1742--1755}.
\bibitem[{Huang and Xiao(2015)}]{huang2015geographic}
\bibinfo{author}{Huang, Q.}, \bibinfo{author}{Xiao, Y.}, \bibinfo{year}{2015}.
\newblock \bibinfo{title}{Geographic situational awareness: mining tweets for disaster preparedness, emergency response, impact, and recovery}.
\newblock \bibinfo{journal}{ISPRS international journal of geo-information} \bibinfo{volume}{4}, \bibinfo{pages}{1549--1568}.
\bibitem[{{Inspector-General for Emergency Management}(2017)}]{inspector2017review}
\bibinfo{author}{{Inspector-General for Emergency Management}}, \bibinfo{year}{2017}.
\newblock \bibinfo{title}{Review of response to the thunderstorm asthma event of 21--22 november 2016. final report}.
\bibitem[{{International Organization for Standardization}(2018)}]{iso22320_2018}
\bibinfo{author}{{International Organization for Standardization}}, \bibinfo{year}{2018}.
\newblock \bibinfo{title}{Security and resilience — emergency management — guidelines for incident management}.
\newblock \bibinfo{howpublished}{International Standard}.
\newblock \bibinfo{note}{{https://www.iso.org/standard/67851.html}}.
\bibitem[{Kamari and Ham(2022)}]{kamari2022ai}
\bibinfo{author}{Kamari, M.}, \bibinfo{author}{Ham, Y.}, \bibinfo{year}{2022}.
\newblock \bibinfo{title}{{AI}-based risk assessment for construction site disaster preparedness through deep learning-based digital twinning}.
\newblock \bibinfo{journal}{Automation in Construction} \bibinfo{volume}{134}, \bibinfo{pages}{104091}.
\bibitem[{Laurila-Pant et~al.(2023)Laurila-Pant, Pihlajam{\"a}ki, Lanki and Lehikoinen}]{laurila2023protocol}
\bibinfo{author}{Laurila-Pant, M.}, \bibinfo{author}{Pihlajam{\"a}ki, M.}, \bibinfo{author}{Lanki, A.}, \bibinfo{author}{Lehikoinen, A.}, \bibinfo{year}{2023}.
\newblock \bibinfo{title}{A protocol for analysing the role of shared situational awareness and decision-making in cooperative disaster simulations}.
\newblock \bibinfo{journal}{International Journal of Disaster Risk Reduction} \bibinfo{volume}{86}, \bibinfo{pages}{103544}.
\bibitem[{Li and Mostafavi(2022)}]{li2022location}
\bibinfo{author}{Li, B.}, \bibinfo{author}{Mostafavi, A.}, \bibinfo{year}{2022}.
\newblock \bibinfo{title}{Location intelligence reveals the extent, timing, and spatial variation of hurricane preparedness}.
\newblock \bibinfo{journal}{Scientific reports} \bibinfo{volume}{12}, \bibinfo{pages}{16121}.
\bibitem[{{Meta}(2025)}]{Meta_Disaster_Maps_2025}
\bibinfo{author}{{Meta}}, \bibinfo{year}{2025}.
\newblock \bibinfo{title}{Facebook disaster maps}.
\newblock \bibinfo{howpublished}{Online}.
\newblock \URLprefix \url{https://dataforgood.facebook.com/dfg/tools/disaster-maps}. \bibinfo{note}{meta Data for Good: Disaster Maps tool for aggregated mobility and displacement insights during crises}.
\bibitem[{Middleton et~al.(2013)Middleton, Middleton and Modafferi}]{middleton2013real}
\bibinfo{author}{Middleton, S.E.}, \bibinfo{author}{Middleton, L.}, \bibinfo{author}{Modafferi, S.}, \bibinfo{year}{2013}.
\newblock \bibinfo{title}{Real-time crisis mapping of natural disasters using social media}.
\newblock \bibinfo{journal}{IEEE Intelligent systems} \bibinfo{volume}{29}, \bibinfo{pages}{9--17}.
\bibitem[{Militello et~al.(2005)Militello, Patterson, Wears and Ritter}]{militello2005large}
\bibinfo{author}{Militello, L.G.}, \bibinfo{author}{Patterson, E.S.}, \bibinfo{author}{Wears, R.}, \bibinfo{author}{Ritter, J.A.}, \bibinfo{year}{2005}.
\newblock \bibinfo{title}{Large-scale coordination in emergency response}, in: \bibinfo{booktitle}{Proceedings of the human factors and ergonomics society annual meeting}, \bibinfo{organization}{SAGE Publications Sage CA: Los Angeles, CA}. pp. \bibinfo{pages}{534--538}.
\bibitem[{Nafday(2009)}]{nafday2009strategies}
\bibinfo{author}{Nafday, A.M.}, \bibinfo{year}{2009}.
\newblock \bibinfo{title}{Strategies for managing the consequences of black swan events}.
\newblock \bibinfo{journal}{Leadership and Management in Engineering} \bibinfo{volume}{9}, \bibinfo{pages}{191--197}.
\bibitem[{{National Institute of Standards and Technology (NIST)}(2015)}]{nist2015_disaster_resilience_framework}
\bibinfo{author}{{National Institute of Standards and Technology (NIST)}}, \bibinfo{year}{2015}.
\newblock \bibinfo{title}{Disaster Resilience Framework}.
\newblock \bibinfo{type}{Technical Report}. National Institute of Standards and Technology (NIST). \bibinfo{address}{Gaithersburg, MD}.
\newblock \URLprefix \url{https://www.nist.gov/system/files/documents/el/building_materials/resilience/Framework_LineNumbered_75-25_11Feb2015.pdf}.
\bibitem[{{Oak Ridge National Laboratory}(2014)}]{ORNL_EARSS_Operational}
\bibinfo{author}{{Oak Ridge National Laboratory}}, \bibinfo{year}{2014}.
\newblock \bibinfo{title}{EARSS Operational Document}.
\newblock \bibinfo{type}{Technical Manual} \bibinfo{number}{Oak Ridge, Tennessee 37831-6007}. Oak Ridge National Laboratory. \bibinfo{address}{Oak Ridge, TN, USA}.
\newblock \URLprefix \url{https://www.ornl.gov/sites/default/files/EARSS-Operational-Document.pdf}.
\bibitem[{O'Brien et~al.(2020)O'Brien, Read and Salmon}]{o2020situation}
\bibinfo{author}{O'Brien, A.}, \bibinfo{author}{Read, G.J.}, \bibinfo{author}{Salmon, P.M.}, \bibinfo{year}{2020}.
\newblock \bibinfo{title}{Situation awareness in multi-agency emergency response: Models, methods and applications}.
\newblock \bibinfo{journal}{International Journal of Disaster Risk Reduction} \bibinfo{volume}{48}, \bibinfo{pages}{101634}.
\bibitem[{Omitaomu et~al.(2010)Omitaomu, Fernandez and Bhaduri}]{omitaomu2010framework}
\bibinfo{author}{Omitaomu, O.A.}, \bibinfo{author}{Fernandez, S.J.}, \bibinfo{author}{Bhaduri, B.L.}, \bibinfo{year}{2010}.
\newblock \bibinfo{title}{Framework for real-time, all-hazards global situational awareness}.
\newblock \bibinfo{journal}{Handbook of Emergency Response} .
\bibitem[{{One Concern, Inc.}(2025)}]{OneConcern_ResiliencePlatform_2025}
\bibinfo{author}{{One Concern, Inc.}}, \bibinfo{year}{2025}.
\newblock \bibinfo{title}{One concern – planetary‑scale resilience software platform}.
\newblock \bibinfo{howpublished}{Online}.
\newblock \URLprefix \url{https://www.oneconcern.com/en/}.
\bibitem[{Pak and Paal(2025)}]{pak2025real}
\bibinfo{author}{Pak, H.}, \bibinfo{author}{Paal, S.G.}, \bibinfo{year}{2025}.
\newblock \bibinfo{title}{A real-time structural seismic response prediction framework based on transfer learning and unsupervised learning}.
\newblock \bibinfo{journal}{Engineering Structures} \bibinfo{volume}{323}, \bibinfo{pages}{119227}.
\bibitem[{{Resilitix Intelligence}(2025)}]{ResilitixAI_2025}
\bibinfo{author}{{Resilitix Intelligence}}, \bibinfo{year}{2025}.
\newblock \bibinfo{title}{Resilitix.ai – ai‑powered disaster resilience platform}.
\newblock \bibinfo{howpublished}{Online}.
\newblock \URLprefix \url{https://www.resilitix.ai/}. \bibinfo{note}{describes Resilitix’s EmergenCITY platform delivering near‑real‑time situational awareness and AI analytics for disaster resilience and emergency management}.
\bibitem[{Richards(2020)}]{richards2020boyd}
\bibinfo{author}{Richards, C.}, \bibinfo{year}{2020}.
\newblock \bibinfo{title}{Boyd’s ooda loop}.
\newblock \bibinfo{journal}{Necesse} \bibinfo{volume}{5}, \bibinfo{pages}{142--165}.
\bibitem[{Riddell et~al.(2019)Riddell, van Delden, Maier and Zecchin}]{riddell2019exploratory}
\bibinfo{author}{Riddell, G.A.}, \bibinfo{author}{van Delden, H.}, \bibinfo{author}{Maier, H.R.}, \bibinfo{author}{Zecchin, A.C.}, \bibinfo{year}{2019}.
\newblock \bibinfo{title}{Exploratory scenario analysis for disaster risk reduction: Considering alternative pathways in disaster risk assessment}.
\newblock \bibinfo{journal}{International journal of disaster risk reduction} \bibinfo{volume}{39}, \bibinfo{pages}{101230}.
\bibitem[{{SafeGraph Inc.}(2024)}]{safegraph2024}
\bibinfo{author}{{SafeGraph Inc.}}, \bibinfo{year}{2024}.
\newblock \bibinfo{title}{Safegraph – the source of truth about physical places}.
\newblock \URLprefix \url{https://www.safegraph.com/}. \bibinfo{note}{accessed: 2025-07-31}.
\bibitem[{Sanz~Segovia(2025)}]{Mapfre_DronesSatellites_2025}
\bibinfo{author}{Sanz~Segovia, G.}, \bibinfo{year}{2025}.
\newblock \bibinfo{title}{Drones and satellites: assessment of damages in affected areas}.
\newblock \bibinfo{journal}{MAPFRE Global Risks – Risks Management and Insurance Magazine} \URLprefix \url{https://www.mapfreglobalrisks.com/en/risks-insurance-management/article/drones-satellites-damage-natural-disasters/}. \bibinfo{note}{accessed 2025}.
\bibitem[{Sepp{\"a}nen and Virrantaus(2015)}]{seppanen2015shared}
\bibinfo{author}{Sepp{\"a}nen, H.}, \bibinfo{author}{Virrantaus, K.}, \bibinfo{year}{2015}.
\newblock \bibinfo{title}{Shared situational awareness and information quality in disaster management}.
\newblock \bibinfo{journal}{Safety Science} \bibinfo{volume}{77}, \bibinfo{pages}{112--122}.
\bibitem[{Songsore(2017)}]{songsore2017complex}
\bibinfo{author}{Songsore, J.}, \bibinfo{year}{2017}.
\newblock \bibinfo{title}{The complex interplay between everyday risks and disaster risks: the case of the 2014 cholera pandemic and 2015 flood disaster in accra, ghana}.
\newblock \bibinfo{journal}{International journal of disaster risk reduction} \bibinfo{volume}{26}, \bibinfo{pages}{43--50}.
\bibitem[{Stanton(2016)}]{stanton2016distributed}
\bibinfo{author}{Stanton, N.A.}, \bibinfo{year}{2016}.
\newblock \bibinfo{title}{Distributed situation awareness}.
\bibitem[{Stanton et~al.(2006)Stanton, Stewart, Harris, Houghton, Baber, McMaster, Salmon, Hoyle, Walker, Young et~al.}]{stanton2006distributed}
\bibinfo{author}{Stanton, N.A.}, \bibinfo{author}{Stewart, R.}, \bibinfo{author}{Harris, D.}, \bibinfo{author}{Houghton, R.J.}, \bibinfo{author}{Baber, C.}, \bibinfo{author}{McMaster, R.}, \bibinfo{author}{Salmon, P.}, \bibinfo{author}{Hoyle, G.}, \bibinfo{author}{Walker, G.}, \bibinfo{author}{Young, M.S.}, et~al., \bibinfo{year}{2006}.
\newblock \bibinfo{title}{Distributed situation awareness in dynamic systems: theoretical development and application of an ergonomics methodology}.
\newblock \bibinfo{journal}{Ergonomics} \bibinfo{volume}{49}, \bibinfo{pages}{1288--1311}.
\bibitem[{{United Nations Office for Disaster Risk Reduction}(2023)}]{united2023global}
\bibinfo{author}{{United Nations Office for Disaster Risk Reduction}}, \bibinfo{year}{2023}.
\newblock \bibinfo{title}{Global Status of Multi-Hazard Early Warning Systems 2023}.
\newblock \bibinfo{publisher}{UN}.
\bibitem[{{US Department of Health \& Human Services}(2025)}]{HHS_emPOWER_program}
\bibinfo{author}{{US Department of Health \& Human Services}}, \bibinfo{year}{2025}.
\newblock \bibinfo{title}{{HHS} empower program}.
\newblock \bibinfo{howpublished}{Online}.
\newblock \URLprefix \url{https://empowerprogram.hhs.gov/}. \bibinfo{note}{accessed via official website including Map, REST Service, AI tools, and datasets}.
\bibitem[{{US Department of Homeland Security Science and Technology Directorate}(2024)}]{US2024HomelandSecurity}
\bibinfo{author}{{US Department of Homeland Security Science and Technology Directorate}}, \bibinfo{year}{2024}.
\newblock \bibinfo{title}{Community Lifeline Status system}.
\newblock \URLprefix \url{https://www.dhs.gov/sites/default/files/202501/25_0127_community_lifeline_status_system_fact_sheet.pdf}.
\bibitem[{{WIFIRE Cyberinfrastructure}(2025)}]{WIFIRE_Firemap_2025}
\bibinfo{author}{{WIFIRE Cyberinfrastructure}}, \bibinfo{year}{2025}.
\newblock \bibinfo{title}{Firemap: Real‑time wildfire behavior modeling and visualization tool}.
\newblock \bibinfo{howpublished}{Online}.
\newblock \URLprefix \url{https://wifire.ucsd.edu/firemap}.
\bibitem[{{World Food Programme (WFP)}(2025)}]{WFP_SKAI_Project_2025}
\bibinfo{author}{{World Food Programme (WFP)}}, \bibinfo{year}{2025}.
\newblock \bibinfo{title}{Skai: Open‑source ai and satellite imagery tool for rapid post‑disaster building damage assessment}.
\newblock \bibinfo{howpublished}{Online}.
\newblock \URLprefix \url{https://innovation.wfp.org/project/SKAI}.
\bibitem[{Yan et~al.(2025)Yan, Yang, Shen, Mahmud, Foutz and Anton}]{yan2025mobirescue}
\bibinfo{author}{Yan, L.}, \bibinfo{author}{Yang, B.}, \bibinfo{author}{Shen, H.}, \bibinfo{author}{Mahmud, S.}, \bibinfo{author}{Foutz, N.Z.}, \bibinfo{author}{Anton, J.}, \bibinfo{year}{2025}.
\newblock \bibinfo{title}{Mobirescue: Optimal dispatching of rescue teams under flooding disasters}.
\newblock \bibinfo{journal}{IEEE Transactions on Mobile Computing} .
\bibitem[{Yuan et~al.(2022)Yuan, Fan, Farahmand, Coleman, Esmalian, Lee, Patrascu, Zhang, Dong and Mostafavi}]{yuan2022smart}
\bibinfo{author}{Yuan, F.}, \bibinfo{author}{Fan, C.}, \bibinfo{author}{Farahmand, H.}, \bibinfo{author}{Coleman, N.}, \bibinfo{author}{Esmalian, A.}, \bibinfo{author}{Lee, C.C.}, \bibinfo{author}{Patrascu, F.I.}, \bibinfo{author}{Zhang, C.}, \bibinfo{author}{Dong, S.}, \bibinfo{author}{Mostafavi, A.}, \bibinfo{year}{2022}.
\newblock \bibinfo{title}{Smart flood resilience: harnessing community-scale big data for predictive flood risk monitoring, rapid impact assessment, and situational awareness}.
\newblock \bibinfo{journal}{Environmental Research: Infrastructure and Sustainability} \bibinfo{volume}{2}, \bibinfo{pages}{025006}.
\bibitem[{{ZestyAI, Inc.}(2025)}]{ZestyAI_ZFIRE_2025}
\bibinfo{author}{{ZestyAI, Inc.}}, \bibinfo{year}{2025}.
\newblock \bibinfo{title}{Z‑fire™: Property‑level wildfire risk scoring model}.
\newblock \bibinfo{howpublished}{Online}.
\newblock \URLprefix \url{https://zesty.ai/products/wildfire}.

\end{thebibliography}

\end{document}